\documentclass{article}
\usepackage[bookmarksnumbered,plainpages]{hyperref}
\usepackage{amsmath}
\usepackage{amssymb}

\newtheorem{theorem}{Theorem}

\def\bt{\begin{tabular}}
\def\et{\end{tabular}}
\def\bc{\begin{center}}
\def\ec{\end{center}}

\def\comb#1#2#3{{\mathsurround 0pt\hbox to 0pt {\hspace*{#3}\raisebox{#2}{${#1}$}\hss}}}
\def\combs#1#2#3{{\mathsurround 0pt\hbox to 0pt {\hspace*{#3}\raisebox{#2}{${\scriptstyle #1}$}\hss}}}
\def\combss#1#2#3{{\mathsurround 0pt\hbox to 0pt {\hspace*{#3}\raisebox{#2}{${\scriptscriptstyle #1}$}\hss}}}

\def\p{\partial}
\def\dfrac#1#2{{\displaystyle\frac{#1}{#2}}}
\def\non{\nonumber}
\def\df{{\rm d}}
\def\eps{\varepsilon}

\def\bfgr#1{\boldsymbol{#1}}
\def\ir#1{\textrm{{\rm #1}}}
\def\bcdot{\mathbin{\bfgr{\cdot}}}
\def\bwedge{\mathbin{\bfgr{\wedge}}}
\def\const{{\rm const}}

\def\bfCl{\mathbf{C}}
\def\bp{\bfgr{\p}}
\def\bim{\bfgr{\imath}}
\def\baab{{\bf b}}
\def\bbaab{{\mathsurround 0pt\mbox{${\bf b}$\hspace*{-1.1ex}${\bf b}$}}}
\def\bFem{{\bf F}}
\def\bEem{{\bf E}}
\def\bBem{{\bf B}}

\def\bjem{{\bf j}}
\def\je{{\mathsurround 0pt\lower.0ex\hbox{${\scriptscriptstyle e}$}\mspace{-4.5mu}j}}
\def\jm{{\mathsurround 0pt\lower.0ex\hbox{${\scriptscriptstyle m}$}\mspace{-7.0mu}j}}
\def\bje{{\mathsurround 0pt\lower.0ex\hbox{${\scriptscriptstyle \mathbf{e}}$}\mspace{-3.4mu}\mathbf{j}}}
\def\bjm{{\mathsurround 0pt\lower.0ex\hbox{${\scriptscriptstyle \mathbf{m}}$}\mspace{-5.6mu}\mathbf{j}}}
\def\Ce{\combss{e}{0.45ex}{0.2em}C}
\def\Cm{\combss{m}{0.45ex}{0.1em}C}
\def\x{{\mathtt{x}}}
\def\z{{\mathtt{z}}}
\def\Sou{S}
\def\Soub{{\bf S}}
\def\Vol{\combs{\square}{0.15ex}{0.2ex} {\rm V}}
\def\Voli{{\combss{\square}{0.12ex}{0.095ex} {\rm V}}}
\def\dVol{{\rm d}\hspace{-0.3ex}\Vol}
\def\Vols{\combs{\triangle}{0.45ex}{0.1ex} {\rm V}}
\def\Volsi{\combss{\triangle}{0.31ex}{-0.02ex} {\rm V}}
\def\dVols{{\rm d}\hspace{-0.3ex}\Vols}
\def\delf{{\mathsurround=0ex {\displaystyle\comb{\cdot}{-0.1ex}{0.07em}\delta}}}
\def\Sur{\Sigma}
\def\Suri{\Sigma}
\def\dSur{{\rm d}\hspace{-0.3ex}\Sigma}
\def\bdSur{{\bf d}\hspace{-0.3ex}{\bfgr{\Sigma}}}
\def\bSur{\Sigma^\prime}
\def\bSuri{\Sigma^\prime}
\def\bdbSur{{\bf d}\hspace{-0.3ex}\bfgr{\Sigma^\prime}}
\def\metr{g}
\def\unitc{{\bf 1}}
\def\gx{\mathfrak{x}}
\def\bfgx{\boldsymbol{\mathfrak{x}}}
\def\tr{\tilde{\gx}}
\def\bftr{\boldsymbol{\tr}}
\def\trj{\mathfrak{t}}
\def\bftrj{\boldsymbol{\trj}}
\def\ra{\acute{\gx}}
\def\bfra{{\boldsymbol{\ra}}}
\def\V{\mathfrak{u}}
\def\bfV{\boldsymbol{\V}}
\def\tadv{t_{{\rm a}}}
\def\tret{t_{{\rm r}}}
\def\dotV{\dot{\V}}
\def\bfdotV{\boldsymbol{\dotV}}

\begin{document}

\title{\bf Source function and dyon's field
in Clifford number representation for electrodynamics}
\author{{\bf Alexander A. Chernitskii}\\
\small St.-Petersburg Electrotechnical University\\
\small  Prof. Popov str. 5, St.-Petersburg, 197376, Russia\\
\small  aa@cher.etu.spb.ru}
\date{}
\maketitle

\begin{abstract}
Clifford number representation for linear electrodynamics with
dyon sources is considered. Source function for the appropriate
system of the first order equations for electromagnetic field is
obtained. The field of an arbitrary moving point dyon is derived.
\end{abstract}

\section{Introduction}

Electrodynamics source function is commonly obtained for
the system of second order wave equations for electromagnetic four-potential.
In some cases this approach is not convenient, in particular,
if there is a source with both electric and magnetic charges
or current densities.
A Clifford number representation for Maxwell system of the first order
differential equations \cite{Riesz1993,Hestenes1984,Ibiclif}
seems to be the most convenient for this case.

\section{Maxwell equations in Clifford numbers}

A Clifford number for space-time can be represented as
\begin{multline}
\bfCl = a\,\unitc {}+{}  c_\mu\,\baab^{\mu}  {}+{}
d_{\mu\nu}\,\baab^{\mu} \bwedge \baab^{\nu} \\ {}+{}
f_{\mu\nu\rho}\,\baab^{\mu} \bwedge  \baab^{\nu} \bwedge   \baab^{\rho}
 {}+{}
q_{\mu\nu\rho\sigma}\,\baab^{\mu} \bwedge  \baab^{\nu} \bwedge \baab^{\rho}
\bwedge \baab^{\sigma}
\quad,
\label{CliffNum}
\end{multline}
where Greek indices take values $0,1,2,3$ and
$\baab^\mu$ are considered as basis vectors in space-time such that
\begin{align}
\baab^\mu \bcdot \baab^\nu &= \unitc\,\metr^{\mu\nu}
\quad,\qquad
\baab^\mu \bcdot \baab_\nu = \unitc\,\metr^\mu_\nu = \unitc\,\delta^\mu_\nu
\quad,\qquad
\baab_\mu \bcdot \baab_\nu = \unitc\,\metr_{\mu\nu}
\quad.
\label{baabbcdotbaab}
\end{align}
Here wedge $\bwedge$ denotes the exterior or antisymmetric product in
Clifford algebra
\begin{equation}
\label{Def:bwedge}
\baab^{\mu} \bwedge \baab^{\nu} \equiv
\frac{1}{2}\left(\baab^{\mu} \, \baab^{\nu} - \baab^{\nu} \, \baab^{\mu}\right)
\quad\phantom{.}
\end{equation}
and dot $\bcdot$ denotes the internal or symmetric one (see also \cite{Ibiclif})
\begin{equation}
\label{Def:bcdot}
\baab^{\mu} \bcdot \baab^{\nu} \equiv
\frac{1}{2}\left(\baab^{\mu} \, \baab^{\nu} + \baab^{\nu} \, \baab^{\mu}\right)
\quad.
\end{equation}

For convenience main designations are collected in appendix.

As it is known the basis vectors $\baab^\mu$ can be represented with Dirac
gamma matrices (see also \cite{Ibiclif}). But in the present article I do not
use any representation for the basis vectors because all results do not need that.

Let us consider that $\bfCl$ in (\ref{CliffNum}) is an invariant
geometrical object, {\rm i.e.} it is an invariant under transformations for
the basis in space-time. In this case
the last term in (\ref{CliffNum}) can be represented in the form
\begin{equation}
\label{Pseudoscalar}
q_{\mu\nu\rho\sigma}\,\baab^{\mu} \bwedge  \baab^{\nu} \bwedge \baab^{\rho}
\bwedge \baab^{\sigma}
 = q^\prime\,\frac{1}{4!}\,\eps_{\mu\nu\rho\sigma}\,
\baab^{\mu} \bwedge  \baab^{\nu} \bwedge \baab^{\rho} \bwedge \baab^{\sigma}
\quad,
\end{equation}
where $\eps_{\mu\nu\rho\sigma}$ are components of fully antisymmetric
unit four-rank tensor or pseudoscalar in space-time such that
$\eps_{0123} = \sqrt{|\metr|}$, $\eps^{0123} = -\sqrt{|\metr|}^{-1}$
for some determined orientation
of coordinate system, $\metr$ is a determinant of a metric tensor
{$\metr\equiv\det (\metr_{\mu\nu})$},
$q^\prime$ is a scalar.

Let us introduce the following designation
\begin{equation}
\bim \equiv  \frac{1}{4!}\,
\eps_{\mu\nu\sigma\rho}\baab^\mu\,\baab^\nu\,\baab^\sigma\,
\baab^\rho = \frac{1}{4!}\,
\eps_{\mu\nu\rho\sigma}\,
\baab^{\mu} \bwedge  \baab^{\nu} \bwedge \baab^{\rho} \bwedge \baab^{\sigma}
\quad.
\label{Def:bim}
\end{equation}

Thus the last term of (\ref{CliffNum}) is represented as $q^\prime\,\bim$.
The invariant geometrical object $\bim$ is a coordinate-free form of the
unit pseudoscalar. It can be called also Clifford imaginary unit because
$\bim^2 = -\unitc$ for any coordinate systems.
The use of $\bim$ in the Clifford number representation for electromagnetic
field, instead of
$\bim^\prime = \baab^0\bwedge\baab^1\bwedge\baab^2\bwedge\baab^3$,
is more convenient for general case, in particular, for curvilinear
coordinates or curved space (see my article \cite{Ibiclif}).
It is evident that $(\bim^\prime)^2 = 1/\metr$ (for space-time
{$\metr < 0$} and {$\metr = -1$} for Cartesian coordinates).
Consequently, $\bim^\prime$
can not be called unit pseudoscalar in general case.
If a direction of any basis vector is changed to opposite one,
for example $\baab^1 \to -\baab^1$, then $\bim^\prime \to -\bim^\prime$
but $\bim$ is not changed because also
$\eps_{\mu\nu\sigma\rho} \to -\eps_{\mu\nu\sigma\rho}$.
There is the relation {$\bim = \pm\sqrt{|\metr|}\,\bim^\prime$}.

It is evident that both
\begin{equation}
\baab^\mu\,\bim = - \bim\,\baab^\mu
\label{CommutVecI}
\end{equation}
and {$\baab^\mu\,\bim^\prime = - \bim^\prime\,\baab^\mu$}
for any coordinate system.

Let us using also the following designations for basis bivectors:
\begin{align}
\nonumber
\bbaab^i \equiv \baab^i {}\bwedge{} \baab^0
&\quad\Longrightarrow\quad
\bbaab^i \bcdot \bbaab^j =
-\metr^{00}\,\metr^{ij} {}+{} \metr^{i0}\,\metr^{j0}
\quad,
\\
\bbaab^i \bcdot \bbaab_j \equiv \unitc\,\delta^i_j
&\quad\Longrightarrow\quad
\bbaab_i = \baab_0 {}\bwedge{} \baab_i
\quad,
\nonumber
\\
&\quad\Longrightarrow\quad
\bbaab^i  =
\left(-\metr^{00}\,\metr^{ij} {}+{} \metr^{i0}\,\metr^{j0}\right)
\bbaab_j
\quad,
\label{Def:bb}
\end{align}
where Latin indices take values $1,2,3$.

Using (\ref{Def:bim}) and (\ref{Def:bb}) we have
\begin{align}
&\baab^i {}\bwedge{} \baab^j = \eps^{0ijl}\,\bim\,\bbaab_l
\quad,\qquad
\baab_i {}\bwedge{} \baab_j = -\eps_{0ijl}\,\bim\,\bbaab^l
\quad,
\label{biwbj}
\end{align}
\begin{align}
&\baab^\mu {}\bwedge{} \baab^\nu {}\bwedge{} \baab^\rho
= -\eps^{\mu\nu\rho\sigma}\,\bim\,\baab_\sigma
\quad.
\label{pvec}
\end{align}

Thus expression (\ref{CliffNum}) can be rewritten in the following form:
\begin{equation}
\bfCl = a\,\unitc {}+{}  c_\mu\,\baab^{\mu}  {}+{}
d^\prime_i\,\bbaab^i {}+{} d_{\prime\prime}^i\,\bim\bbaab_i  {}+{}
f_\prime^\mu\,\bim\baab_{\mu}  {}+{} q^\prime\,\bim
\quad.
\label{CliffNumWithI}
\end{equation}
Let us use the following appellations for the terms into
(\ref{CliffNumWithI}): the first term is real scalar, the second one is
real vector, the third one is real bivector, the fourth one is
imaginary bivector, the fifth one is imaginary vector, and
the sixth one is imaginary scalar.

Let us write also the following useful relations:
\begin{align}
\nonumber
&\begin{array}{l}
\baab^0 {}\cdot{} \bbaab_i {}={}  \baab_0 {}\cdot{} \bbaab^i =
\baab^0 {}\cdot{} \bbaab^i {}={}  \baab_0 {}\cdot{} \bbaab_i = 0
\quad,
\end{array}
\\[1.5ex]
&\begin{array}{lll}
\baab^i {}\wedge{} \bbaab_j = -\delta^i_j\,\baab_0
\;,&\baab^0 {}\wedge{} \bbaab_i = \phantom{-}\baab_i\;,&
\baab^i {}\cdot{} \bbaab^j = -\varepsilon^{0ijk}\,\bim\,\baab_k
\;,
\\[1ex]
\baab_i {}\wedge{} \bbaab^j = \phantom{-}\delta_i^j\,\baab^0
\;,&\baab_0 {}\wedge{} \bbaab^i = -\baab^i\;,&
\baab_i {}\cdot{} \bbaab_j = \phantom{-}\varepsilon_{0ijk}\,\bim\,\baab^k
\;.
\end{array}
\label{Tools}
\end{align}

Now let us write the Clifford number form for Maxwell system
of equations \cite{Riesz1993,Hestenes1984,Ibiclif} with both electric
and magnetic four-current densities:
\begin{align}
\bp\,\bFem &= -4\pi\,\bjem
\quad,
\label{MaxwellEq}
\end{align}
where
\begin{align}
&\bFem = \frac{1}{2}\,F_{\mu\nu}\,\baab^{\mu}\,\baab^{\nu} {\,}={\,}
E_i\,\bbaab^{i} {}-{} B^i\,\bim\bbaab_{i}
= \bEem  {}-{} \bim\,\bBem
\quad,
\label{Def:bFem}
\\
&\bp \equiv  \baab^{\mu}\,\p_\mu {\,}\equiv{\,}
\baab^{\mu}\,\dfrac{\p}{\p x^\mu}
\quad,
\label{defbp}
\\
&\bjem \equiv \je_\mu\,\baab^{\mu} {}+{} \jm_\mu\,\bim\baab^{\mu}
= \bje {}+{} \bim\,\bjm
\quad,
\label{Def:bjem}
\end{align}
where $\bje$ and $\bjm$ are electric and magnetic four-current
density, $\je_\mu$ and $\jm_\mu$ are their components.

\section{Source function}

\index{electrodynamics!source function}%

Let us designate space-time points by $\x$, $\z$, where
$\x = \{x^0,x^1,x^2,x^3\}$, $\z = \{z^0,z^1,z^2,z^3\}$.

Let us use the following definition for four-dimensional Dirac delta symbolic
function in any coordinates
(such three-dimensional definition is used in book \cite{Landau&LifschitzII}):
\begin{equation}
\label{Def:delfint}
f(\z) =
\int\limits_{\Voli} f(\x-\z)\,\delf^4 (\x)\,\df x^0 \df x^1 \df x^2 \df x^3
= \int\limits_{\Voli} f(\x-\z)\,\dfrac{\delf^4 (\x)}{\sqrt{|\metr|}}\,\dVol
\quad,
\end{equation}
where $\dVol = \sqrt{|\metr|}\,\df x^0 \df x^1 \df x^2 \df x^3$
is a four-dimensional volume element
and $\Vol$ is a four-volume including the point $\z$.
Here $f(\x)$ is a function which is continuous at the point $\z$.
If a discontinuous function at the point
$\z$ is considered then definition   (\ref{Def:delfint}) must be modified
(see \cite{Chernitskii1998a}).

Thus the following definition is considered for any coordinates
\begin{align}
\delf^4(\x) &= \delf(x^0)\,\delf(x^1)\,\delf(x^2)\,\delf(x^3)
\quad,
\label{FourDelta}
\end{align}
where $\delf(x^\mu)$ is one-dimensional delta symbolic function
such that
\begin{equation}\label{Def:delfint1}
\phi(t) = \int\limits_{l_1}^{l_2}
 \phi(t^\prime - t)\,\delf (t^\prime)\,\df t^\prime
\quad,
\end{equation}
where {$l_1 < t < l_2$}.

Let us designate the three-dimensional hypersurface bounding a four-volume
$\Vol$ as $\Sur$ and its inside oriented element as
{$\bdSur \equiv \dSur_\mu\,\baab^\mu$} such that
\begin{equation}
\begin{array}{l}
\dSur_0 \equiv \pm\sqrt{|\metr|}\,\df x^1 \df x^2 \df x^3
\quad, \\
\dSur_1 \equiv\pm\sqrt{|\metr|}\,\df x^0 \df x^2 \df x^3
\quad, \\
\dSur_2 \equiv \pm\sqrt{|\metr|}\,\df x^0 \df x^1 \df x^3
\quad, \\
\dSur_3 \equiv \pm\sqrt{|\metr|}\,\df x^0 \df x^1 \df x^2
\quad.
\end{array}
\label{Def:bdSur}
\end{equation}

Now let us formulate the following two simple theorems:
\begin{theorem}
If a vector function {$\Soub\equiv\Sou_\mu\,\baab^\mu$},
$\Soub {}={} \Soub (\x)$ is a particular solution to the equation
\begin{equation}
\bp\,\Soub {}={} -\unitc\,4\pi\,\dfrac{\delf^4(\x)}{\sqrt{|\metr|}}
\quad,
\label{Theor1a}
\end{equation}
then a particular solution to Maxwell equation (\ref{MaxwellEq})
into the volume $\Vol$ is given by formula
\begin{equation}
\bFem (\z) =
\int\limits_{\Voli}
\Bigl[
\Soub (\x {}-{} \z) {}\bwedge{} \bje (\x) {}-{}
\bim\,\Soub (\x {}-{} \z) {}\bwedge{} \bjm (\x)
\Bigr]
\dVol
\quad.
\label{Theor1c}
\end{equation}
\vspace{1.5ex}
\label{Theor1}
\end{theorem}

Equation (\ref{Theor1a}) can be rewrite also in the form without
delta function:
$$
\left\{
\begin{array}{l}
\bp\,\displaystyle\Soub {}={} \mathbf{0} \quad\mbox{for}\quad \x\neq \mathtt{0}
\quad,\\[1.5ex]
\displaystyle\lim_{\bSuri\to 0}\int\limits_{\bSuri}\Soub (\mathtt{0})\,\bdbSur {}={}
4\pi\,\unitc
\quad,
\end{array}
\right.
\eqno{(\ref{Theor1a}^\prime)}
$$
where $\bSur$ is a small hypersurface bounding the point $\x=\mathtt{0}$
and {$\bdbSur$} is its outside oriented surface element.

{\bf Proof.} Let us multiply equation (\ref{MaxwellEq}) by
$\Soub (\x {}-{} \z)$ from the left. Then the integration of the result over
space-time volume $\Vol$ gives
\begin{align}
\int\limits_{\Voli}\Soub (\x {}-{} \z)\,\bp\,\bFem (\x)\,\dVol
&= -4\pi\,\int\limits_{\Voli}\Soub (\x {}-{} \z)\,\bjem (\x)\,\dVol
\quad.
\label{proof1.1}
\end{align}
And the integration by parts of the left-hand side in (\ref{proof1.1})
gives
\begin{multline}
-\int\limits_{\Suri}\Soub (\x {}-{} \z)\,\baab^\mu\,\bFem
(\x)\,\dSur_\mu -\int\limits_{\Voli} \left\{\dfrac{\p}{\p
x^\mu}\left[ \Soub (\x {}-{}
\z)\,\baab^\mu\,\sqrt{|\metr|}\right]\right\}\bFem (\x)\, \df x^0
\df x^1 \df x^2 \df x^3
\\
= -4\pi\,\int\limits_{\Voli}\Soub (\x {}-{} \z)\,\bjem (\x)\,\dVol
\quad.
\label{proof1.2}
\end{multline}

From (\ref{Theor1a}) we have
\begin{equation}
\bp\bcdot\Soub {}={} -\unitc\,4\pi\,\dfrac{\delf^4(\x)}{\sqrt{|\metr|}}
\quad,\qquad \bp\bwedge\Soub {}={} \mathbf{0}
\quad.
\label{proof1.2a}
\end{equation}
Also
\begin{align}
\dfrac{\p}{\p x^\mu}
\left[\baab^\mu\,\sqrt{|\metr|}\right] &= \mathbf{0}
\quad.
\label{proof1.2b}
\end{align}
Because of (\ref{proof1.2a}) and (\ref{proof1.2b}) the
second integral into left-hand side of (\ref{proof1.2}) is equal
to {$4\pi\,\bFem (\z)$}. Thus we have
\begin{align}
\bFem (\z) &= \int\limits_{\Voli} \Soub (\x {}-{} \z) \, \bjem (\x)\,\dVol
 {}-{} \frac{1}{4\pi}\int\limits_{\Suri}\Soub (\x {}-{} \z)\,\bdSur\,\bFem (\x)
\quad.
\label{proof1.3}
\end{align}
The action of operator {$\bp$} to equation (\ref{proof1.3})
 and the use (\ref{Theor1a}) give
\begin{align}
\bp\,\bFem (\z) &= -4\pi\,\int\limits_{\Voli}
\frac{\delf^4(\x - \z)}{\sqrt{|\metr|}}\,\bjem (\x)\,\dVol
 {}+{} \int\limits_{\Suri}
\frac{\delf^4(\x - \z)}{\sqrt{|\metr|}}\,\bdSur\,\bFem (\x)
\quad.
\label{proof1.3a}
\end{align}
If $\z\in\Vol$ and $\z\not\in\Sur$ the first term into the right-hand
side of (\ref{proof1.3a})  equals the right-hand
side of equation (\ref{MaxwellEq})
and the second term into the right-hand
side of (\ref{proof1.3a}) equals {${\bf 0}$} (because of $\z\not\in\Sur$).
If a particular solution to equation (\ref{MaxwellEq}) is searched then
only the first term into the right-hand side of (\ref{proof1.3}) can be considered.
But since electromagnetic field {$\bFem$} is bivector one, we must take the
bivector part of the right-hand side into (\ref{proof1.3}).
Using (\ref{Def:bjem}) and (\ref{CommutVecI}) we finally obtain
(\ref{Theor1c}) as a particular solution to Maxwell equation (\ref{MaxwellEq})
{\bf which was to be proved}.

For the proof we can use also expression
{\mathsurround 0pt (\ref{Theor1a}${}^\prime$)} instead
of expression (\ref{Theor1a}). In this case we must make integration into
(\ref{proof1.1}) over the volume {$\Vol^\prime$} which is
the volume $\Vol$ without the small region
including the point $\x=\z$ and bounded by the small surface {$\bSur$}.
Instead of (\ref{proof1.2}) we obtain
\begin{multline}
-\int\limits_{\Suri}\Soub (\x {}-{} \z)\,\bdSur\,\bFem (\x)
-\int\limits_{\Voli^\prime}\left[\bp\,\Soub (\x {}-{} \z)\right]\bFem (\x)\,\dVol
-\int\limits_{\bSuri}\Soub (\x {}-{} \z)\,\bdbSur\,\bFem (\x)\\
= -4\pi\,\int\limits_{\Voli^\prime}\Soub (\x {}-{} \z)\,\bjem (\x)\,\dVol
\;\;.
\label{proof1.4}
\end{multline}
Contracting the surface $\bSur$ and using
{\mathsurround 0pt (\ref{Theor1a}${}^\prime$)}
we obtain (\ref{proof1.3}).

Let us also write formula (\ref{Theor1c}) in components:
\begin{multline}
\bFem (\z) = \int\limits_{\Voli}
\biggl\{
\Bigl[\Sou_i \, \je_0  {}-{} \Sou_0 \, \je_i
 {}-{} \eps_{0ijl}\,\Sou^j \, \jm^l \Bigr]\bbaab^i
\\
 {}-{} \Bigl[ \Sou^0 \, \jm^i {}-{} \Sou^i \, \jm^0 {}-{}
\eps^{0ijl}\,\Sou_j \, \je_l \Bigr]\bim\bbaab_i
\biggr\}\;\dVol
\quad,
\label{Theor1e}
\end{multline}
where {$\Sou_\mu =\Sou_\mu (\x {}-{} \z)$},
{$\je_\mu = \je_\mu (\x)$},
{$\jm_\mu = \jm_\mu (\x)$},
in general case {$\bbaab = \bbaab (\x)$}.

\begin{theorem}
A relativistic invariant source function for Maxwell equation
(\ref{MaxwellEq}) can be taken in the form
\begin{align}
\Soub&= -\bp\,\delf (x^\mu x_\mu)
\quad.
\label{Theor2a}
\end{align}
\label{Theor2}
\end{theorem}

{\bf Proof.} Let us use an inertial coordinate system for which
$\metr_{00} = -1$, $\metr_{0i} = 0$.
Thus $x^\mu x_\mu = -t^2  {}+{} \gx^2$,
where $t\equiv x^0$, $\gx \equiv \sqrt{x^i x_i}$.
(Let us use Gothic letters for designation three-dimensional quantities.)
Then we can write \cite{DiracPQM1958,Richtmyer1978}
\begin{align}
\delf (x^\mu x_\mu) &= \delf (-t^2  {}+{} \gx^2) = \delf (t^2  {}-{} \gx^2)
= \dfrac{1}{2\,\gx}\left[\delf (t {}-{} \gx) {}+{} \delf (t {}+{} \gx)\right]
\quad.
\label{proof2.1}
\end{align}

Let us substitute (\ref{Theor2a}) into the left-hand side of (\ref{Theor1a}).
Thus we have
\begin{align}
\bp\,\Soub &= -\bp\,\bp\,\delf (x^\mu x_\mu)
= -\unitc\,\p_\nu\p^\nu \,\delf (x^\mu x_\mu)
\quad.
\end{align}
But using (\ref{proof2.1}) we have
\begin{multline}
\p_\nu\p^\nu \,\delf (x^\mu x_\mu) =
\left(-\p_0\p_0 {}+{} \Delta\right)
\dfrac{1}{2\,\gx}\left[\delf (t {}-{} \gx) {}+{} \delf (t {}+{} \gx)\right]\\
= \dfrac{1}{2}\left[\delf (t {}-{} \gx) {}+{} \delf (t {}+{} \gx)\right]
\Delta\dfrac{1}{\gx}
= -\frac{1}{2} \left[\delf (t {}-{} \gx) {}+{} \delf (t {}+{} \gx)\right]
4\pi\,\frac{\delf^3(\bfgx)}{\sqrt{|\metr|}} \\
= -4\pi\,\delf(t)\,\frac{\delf^3(\bfgx)}{\sqrt{|\metr|}} =
-4\pi\frac{\delf^4(\x)}{\sqrt{|\metr|}}
\quad.
\label{proof2.2}
\end{multline}
Thus function (\ref{Theor2a}) satisfies equation (\ref{Theor1a})
and the assertion {\bf is proved}.

Using (\ref{proof2.1}) we can write the source function in the
following expanded form:
\begin{multline}
\Soub = \dfrac{1}{2}\left\{-\dfrac{\baab^0}{\gx}
\left[\delf^\prime (t {}-{} \gx) {}+{} \delf^\prime (t {}+{} \gx)\right]\right.\\
 \left.{}+{} \dfrac{\bfgx}{\gx^3}
\left[\delf (t {}-{} \gx) {}+{} \delf (t {}+{} \gx)\right]
{}+{} \dfrac{\bfgx}{\gx^2}
\left[\delf^\prime (t {}-{} \gx) {}-{} \delf^\prime (t {}+{} \gx)\right]
\vphantom{\frac{\baab^0}{\gx}}\right\}
\quad,
\label{SourceFunInCC}
\end{multline}
where {$\bfgx \equiv x^i\,\baab_i$}.
But we must remember that this expression is obtained for
an inertial coordinate system.

\section{Retarded and advanced fields}

\index{electrodynamics!retarded and advanced fields}%
Realizing the integration over time in formula (\ref{Theor1c})
with expression (\ref{SourceFunInCC}) we obtain
\begin{align}
\nonumber
\bFem (\z) &= \dfrac{1}{2}\int\limits_{\Volsi}\biggl[
\dfrac{\bftr}{\tr^3}\bwedge\left(\bje_\ir{ret}  {}+{} \bje_\ir{adv} \right)
 {}+{}
\bim\,\dfrac{\bftr}{\tr^3}\bwedge\left(\bjm_\ir{ret}  {}+{} \bjm_\ir{adv} \right)
\\
\nonumber
&\quad{}+{}
\left(\dfrac{\baab^0}{\tr}  {}+{} \dfrac{\bftr}{\tr^2}\right)\bwedge
\p_0\,\bje_\ir{ret} {}+{}
\bim\left(\dfrac{\baab^0}{\tr}  {}+{} \dfrac{\bftr}{\tr^2}\right)\bwedge
\p_0\,\bjm_\ir{ret}
\\
&\quad {}+{} \left(\dfrac{\baab^0}{\tr} {}-{} \dfrac{\bftr}{\tr^2}\right)\bwedge
\p_0\,\bje_\ir{adv}
 {}+{} \bim\left(\dfrac{\baab^0}{\tr} {}-{} \dfrac{\bftr}{\tr^2}\right)\bwedge
\p_0\,\bjm_\ir{adv}
\biggr]\dVols
\quad,
\label{RetAdvField}
\end{align}
where $\dVols\equiv\sqrt{|\metr|}\,\df x^1\df x^2\df x^3$ 
is a three-dimensional volume element
and $\Vols$ is a three-volume including the point $\bfgx$, $\bftr\equiv \baab_i\,(x^i - z^i)$,
$\tr \equiv \sqrt{(x^i - z^i)(x_i - z_i)}$, and
for a given function $\bjem = \bjem (x^0,\,x^i)$ we have
$\bjem_\ir{ret} \equiv \bjem (z^0 - \tr,\,x^i)$,
$\bjem_\ir{adv} \equiv \bjem (z^0 + \tr,\,x^i)$.

Here it must be noted that the retarded current {$\bjem_\ir{ret}$}
is caused by convergent singular waves {$\delf (x^0 - z^0 + \tr)/\tr^2$},
{$\delf^\prime (x^0 - z^0 + \tr)/\tr$} in source function
{$\Soub (\x-\z)$} (\ref{SourceFunInCC})
and the advanced current {$\bjem_\ir{adv}$} is caused by
divergent singular waves {$\delf (x^0 - z^0 - \tr)/\tr^2$},
{$\delf^\prime (x^0 - z^0 - \tr)/\tr$}.
This is easily to understand because convergent and divergent waves
are considered here with respect to a point of observation $\z$.

\section{Field of moving dyon}

\index{electrodynamics!dyon's field}%

A current density for moving point dyon in inertial coordinates is
\begin{align}
\bjem &= \frac{1}{\sqrt{|\metr|}}\left(\unitc\,\Ce {}+{} \bim\,\Cm\right)
\left(\baab^0  {}+{} \bfV\right)
\delf^3 \bigl(\bfgx {}-{} \bftrj(t)\bigr)
\quad,
\label{DyonCurrent}
\end{align}
where  $\Ce$ is  an electric charge and $\Cm$ is a magnetic one,
$\bfV = \V_i\,\baab^i$ is a velocity of the dyon,
{$\bftrj(t)= \trj_i(t)\,\baab^i$} is a trajectory of the dyon.
Let us designate
\begin{equation}
\bfra = \ra_i\,\baab^i \equiv \bfgx {}-{} \bftrj(t)
\quad.
\end{equation}
For the velocity we have the following relation:
\begin{align}
\bfV &= \dfrac{\df \bftrj}{\df t} = -\dfrac{\df \bfra}{\df t}
\quad.
\label{Def:Vel}
\end{align}

Now let us substitute dyon singular current (\ref{DyonCurrent}) and
source function (\ref{SourceFunInCC}) into formula (\ref{Theor1c})
and then make the integration over a three-dimensional volume
$\Vols$. As result we have the following expression for dyon's
field:
\begin{multline}
\bFem (t,x^i) = \dfrac{1}{2}\int\limits_{-\infty}^{\infty}
\left\{\left(\unitc\,\Ce {}-{} \bim\,\Cm\right)
\left[\dfrac{\bfV\bwedge\baab^0}{\ra}
\Bigr(\delf^\prime (t^\prime  {}-{} t {}-{}\ra) {}+{}
\delf^\prime (t^\prime  {}-{} t {}+{}\ra)\Bigl)\right.\right.\\
 {}+{}
\left(\bfra\bwedge\baab^0 {}+{} \bfra\bwedge\bfV\right)\biggl(
-\dfrac{1}{\ra^3}
\Bigr(\delf (t^\prime  {}-{} t {}-{}\ra) {}+{}
\delf (t^\prime  {}-{} t {}+{}\ra)\Bigl)\\
 \left.\left.{}+{}
\dfrac{1}{\ra^2}
\Bigr(\delf^\prime (t^\prime  {}-{} t {}-{}\ra) {}-{}
\delf^\prime (t^\prime  {}-{} t {}+{}\ra)\Bigl)\biggr)\right]
\right\}\df t^\prime
\quad,
\label{DyonFieldT}
\end{multline}
where {$\bfra = \bfra (t^\prime)$}, {$\bfV = \bfV (t^\prime)$}.
(Here for the convenience the variables in formula (\ref{Theor1c})
are changed: {$\x \to \x^\prime$}, {$\z \to \x$}.)

Realizing the integration over time into (\ref{DyonFieldT}) we must
take into account that the argument of delta function is
complicated because of {$\ra = \ra (t^\prime)$}.
For this case we have the following relations:
\begin{align}
\int\limits_{l_1}^{l^2} f(t)\,\delf\bigl(\varphi(t)\bigr)\,\df t &=
\sum\limits_s\frac{f(t_s)}{\varphi^\prime (t_s)}
\quad,\\
\int\limits_{l_1}^{l^2} f(t)\,\delf^\prime\bigl(\varphi(t)\bigr)\,\df t &=
\sum\limits_s\left[\frac{f\,\varphi^{\prime\prime}}{(\varphi^\prime)^3}
 {}-{} \frac{f^\prime}{(\varphi^\prime)^2}\right]_{t=t_s}
\quad,
\end{align}
where $t_s$ is a solution of the equation {$\varphi (t) = 0$}
such that {$l_1<t_s<l_2$}.

Using {$\Ce = \const$} and {$\Cm = \const$},
after necessary transformations we obtain
\begin{align}
\non
\bFem (t,x^i) &= \dfrac{1}{2}\,
\left(\unitc\,\Ce {}-{} \bim\,\Cm\right)
\Bigl\{
\\
\non &\phantom{+}\Bigl[\left(\ra {}-{} \ra^i\,\V_i\right)^{-3}
\left(1 {}-{} \V^2 {}+{} \ra^i\,\dotV_i\right)
\left(\bfra\bwedge\baab^0 {}-{} \ra\,\bfV\bwedge\baab^0 {}-{}
\bfra {}\bwedge{} \bfV \right)
\\
\non &\quad -\left(\ra {}-{} \ra^i\,\V_i\right)^{-2}
\Bigl(\ra\,\bfdotV\bwedge\baab^0 {}+{} \bfra {}\bwedge{} \bfdotV
\Bigr) \Bigr]_{{\rm ret}}
\\
\non &+\Bigl[\left(\ra {}+{} \ra^i\,\V_i\right)^{-3} \left(1 {}-{}
\V^2 {}+{} \ra^i\,\dotV_i\right) \left(\bfra\bwedge\baab^0 {}+{}
\ra\,\bfV\bwedge\baab^0 {}-{} \bfra {}\bwedge{} \bfV \right)
\\
&\quad -\left(\ra {}+{} \ra^i\,\V_i\right)^{-2}
\Bigl(\ra\,\bfdotV\bwedge\baab^0 {}-{} \bfra {}\bwedge{} \bfdotV
\Bigr) \Bigr]_{{\rm adv}}\; \Bigr\} \quad,
\label{DyonField}
\end{align}
where for {$\bigl[...\bigr]_{{\rm ret}}$} we make the substitution
{$t=\tret$ (retarded time)} and
for {$\bigl[...\bigr]_{{\rm adv}}$}
we make the substitution
{$t=\tadv$ (advanced time)}. Retarded and advanced times
are obtained from the following equations:
\begin{equation}
\tret {}-{} \left[t {}-{} \ra(\tret)\right] = 0
\quad,\qquad
\tadv {}-{} \left[t {}+{} \ra(\tadv)\right] = 0
\quad,
\end{equation}
{$\bfdotV$} is an acceleration of the dyon such that
\begin{align}
\bfdotV &\equiv \dfrac{\df\bfV}{\df t}
\quad.
\label{Def:accel}
\end{align}

For the case of $\Cm=0$ and if we obliterate the advanced path in
expression (\ref{DyonField}), it gives the customary retarded field of
a moving point electrical charge. A direct verification of
the appropriate filed components gives the complete agreement with
the known classical result for this case \cite{Landau&LifschitzII}.

\section{Conclusions}

Thus I have obtained the source function (\ref{Theor2a}) for
Maxwell equations (with electric and magnetic current densities)
 in Clifford number representation (\ref{MaxwellEq}).
It is essential that the source function has the evident relativistic
invariance that can use also for the case of a noninertial coordinate
system or curved space. This source function
include both divergent and convergent singular waves.

I have obtained the three-dimensional integral representation
for a particular solution of (\ref{MaxwellEq}) which includes
retarded and advanced parts.

I have obtained the dyon's field which includes both divergent and
convergent waves. In some problems such representations for
particular solution of Maxwell equations can be helpful (see
\cite{Idyonint}).

\begin{appendix}
\section{Main designations}

\bc\bt{|c|p{45ex}|c|}
\hline
symbol & name & appearance\\
\hline
\hline $\unitc$ & generalized unit & (\ref{CliffNum})\\
\hline $\baab^{\mu}$ & basis vectors & (\ref{CliffNum}) \\
\hline $\metr_{\mu\nu}$, $\metr$ & metric tensor and its determinant & (\ref{baabbcdotbaab})\\
\hline $\bwedge$ & exterior or antisymmetric product & (\ref{Def:bwedge}) \\
\hline $\bcdot$ & inner or symmetric product & (\ref{Def:bcdot}) \\
\hline $\eps_{\mu\nu\rho\sigma}$ & fully antisymmetric unit four-rank tensor & (\ref{Pseudoscalar})\\
\hline $\bim$ & unit pseudoscalar & (\ref{Def:bim})\\
\hline $\bbaab^{i}$ & basis bivectors & (\ref{Def:bb})\\
\hline $\bFem$  & electromagnetic bivector & (\ref{Def:bFem})\\
\hline $\bp$ & operator of differentiation& (\ref{defbp})\\
\hline $\bjem$  & electromagnetic four-current density & (\ref{Def:bjem})\\
\hline $\bje$, $\je_\mu$, $\bjm$, $\jm_\mu$ & electric and magnetic four-current densities & (\ref{Def:bjem})\\
\hline $\x$, $\z$ & space-time points & (\ref{Def:delfint})\\
\hline $\Vol$, $\dVol$ & four-volume and its element &(\ref{Def:delfint})\\
\hline $\delf$ & Dirac delta symbolic function & (\ref{FourDelta})\\
\hline $\Sur$; $\bdSur$, $\dSur_\mu$& three-dimensional hypersurface; its element & (\ref{Def:bdSur})\\
\hline $\Soub$, $\Sou_\mu$ & source function & (\ref{Theor1a}) \\
\hline $\bfgx$, $\gx_i$ & three-dimensional position vector &(\ref{SourceFunInCC})\\
\hline $\Vols$, $\dVols$ & three-volume and its element &(\ref{RetAdvField})\\
\hline $\Ce$, $\Cm$& electric and magnetic charges of a dyon &(\ref{DyonCurrent})\\
\hline $\bftrj$, $\trj_i$ & trajectory of the dyon &(\ref{DyonCurrent})\\
\hline $\bfV$, $\V_i$&velocity of the dyon &(\ref{DyonCurrent})\\
\hline $\bfdotV$, $\dotV_i$&acceleration of the dyon&(\ref{Def:accel})\\ \hline
\et\ec
\end{appendix}

\begin{thebibliography}{99}
\def\topsep{0pt}
\def\parsep{0pt plus 5pt minus 1pt}
\def\itemsep{-0.5ex} 
\small               

\bibitem{Riesz1993}
 M. Riesz, {\em Clifford Numbers and Spinors},
Fundamental Theories of Physics, vol. 54, Kluwer Academic Publishers,
Dordrecht, 1993.

\bibitem{Hestenes1984}
D. Hestenes and G. Sobczyk,
{\em Clifford algebra to geometric calculus.
A unified language for mathematics and physics},
Fundamental Theories of Physics, vol. 18,
D.~Reidel Publishing, Dordrecht, 1984.

\bibitem{Ibiclif} A.A. Chernitskii, {
``Born-Infeld electrodynamics: Clifford number and spinor
representations''},
{\em Int. J. Math. \& Math. Sci.} {\bf 31} (2002), 77--84;
\mbox{hep-th/0009121}.

\bibitem{Landau&LifschitzII}
L.D. Landau and E.M. Lifshitz, {\em The Classical Theory
             of Fields}, Pergamon Press, Oxford, 1975.

\bibitem{Chernitskii1998a} A.A.~Chernitskii,
``Nonlinear electrodynamics
with singularities (modernized Born-Infeld electrodynamics)'',
{\em Helv. Phys. Acta.} {\bf 71} (1998), 274--287;
\mbox{hep-th/9705075}.

\bibitem{DiracPQM1958} P.A.M. Dirac,
{\em The Principles of quantum mechanics}, fourth edition, Oxford, 1958.

\bibitem{Richtmyer1978} R.D. Richtmyer, {\em Principles of advanced
mathematical physics}, Springer-Verlag, New York, 1978.

\bibitem{Idyonint}
A.A. Chernitskii, ``Dyons and interactions in nonlinear
(Born-Infeld) electrodynamics'', {\em J. High Energy Phys.}
{\bf 1999} (1999), no. 12, Paper 10, 1--34; \mbox{hep-th/9911093}.

\end{thebibliography}
\end{document}